\documentclass[prb,aps,floats,amssymb,showkeys,showpacs,preprint,
bibnotes, tightenlines]{revtex4}
\usepackage{longtable}
\usepackage{graphicx}
\usepackage{graphics}
\usepackage{epsfig,bm}
\usepackage{rotating}
\usepackage{amsmath}
\begin{document}

\title{ Higher order expansions for the entropy of a dimer or a monomer-dimer
system on $d$-dimensional lattices}

\author{P. Butera}
\email{paolo.butera@mib.infn.it}
\affiliation
{Dipartimento di Fisica Universita' di Milano-Bicocca\\
and\\
Istituto Nazionale di Fisica Nucleare \\
Sezione di Milano-Bicocca\\
 3 Piazza della Scienza, 20126 Milano, Italy}
\author{P. Federbush}
\email{pfed@umich.edu}
\affiliation
{Department of Mathematics\\
University of Michigan \\
Ann Arbor, MI 48109-1043, USA\\}
\author{M. Pernici} 
\email{mario.pernici@mi.infn.it}
\affiliation
{Istituto Nazionale di Fisica Nucleare \\
Sezione di Milano\\
 16 Via Celoria, 20133 Milano, Italy}

\date{\today}

\begin{abstract}

  Recently an expansion as a power series in $1/d$ has been presented
  for the specific entropy of a complete dimer covering of a
  $d$-dimensional hypercubic lattice.  This paper extends from 3 to 10
  the number of terms known in the series.  Likewise an expansion for
  the entropy, dependent on the dimer-density $p$, of a monomer-dimer
  system, involving a sum $\sum_k a_k(d) p^k$, has been recently
  offered.  We herein extend the number of the known expansion
  coefficients from 6 to 20 for the hyper-cubic lattices of general
  dimension $d$ and from 6 to 24 for the hyper-cubic lattices of
  dimensions $d < 5 $.  We show that this extension can lead to
  accurate numerical estimates of the $p$-dependent entropy for
  lattices with dimension $d > 2$.  The computations of this paper
  have led us to make the following marvelous conjecture: {\it In the
  case of the hyper-cubic lattices, all the expansion coefficients, $
  a_k(d) $, are positive! } This paper results from a simple melding
  of two disparate research programs: one computing to high orders the
  Mayer series coefficients of a dimer gas, the other studying the
  development of entropy from these coefficients.  An effort is made
  to make this paper self-contained by including a review of the
  earlier works.

\end{abstract}
\maketitle
\section{Introduction and results}
The dimer problem arose in a thermodynamic study of diatomic molecules
and was abstracted into one of the most basic and natural problems in
both statistical mechanics \cite{10, 11,19} 
and combinatorial mathematics\cite{fla}.
In more recent years, dimers found interesting applications also in
information \cite{7} and string theories\cite{IQ2,IQ3}. 

 Given a hyper-simple-cubic (hsc) lattice with number of sites $N$ 
in $d$ dimensions, the dimer problem loosely speaking is to count the
number of different ways dimers (dominoes) may be laid down in the
lattice (without overlapping) to completely cover it. Each dimer
covers two nearest neighbor vertices. It is known\cite{ham} that the number of
such coverings is roughly $\exp (\lambda_d N$) for some constant
$\lambda_d$ as $N$ goes to infinity.  In 1980 H.Minc\cite{1} gave a
proof of the asymptotic relation (asymptotic as $d \to \infty$)
\begin{equation}
\lambda_d  \sim \frac{1} {2} {\rm ln}(2d) - \frac{1} {2}.
\end{equation} 
In a series of papers\cite{2,3,4,14}, one of the authors, P.F.,
found a mathematical argument for a full asymptotic expansion
\begin{equation}
\lambda_d  \sim \frac{1} {2} {\rm ln}(2d) - \frac{1} {2}+ \frac{c_1} {d} + \frac{c_2} {d^2}+\cdots
\label{1sud}
\end{equation} 
and computed the first three terms in the Table \ref{tab1} also making
the conjecture that no further terms would be computed. He was very
wrong! One of the results of the present paper is the set of
coefficients from $c_4$ to $c_{10}$ reported in Table \ref{tab1}.
\begin{table}[ht]
  \caption{ Expansion coefficients $c_n$   of the dimer entropy 
    $\lambda_d  \sim \frac{1} {2} {\rm ln}(2d) - \frac{1} {2}
    + \sum_n\frac{c_n} {d^n}$
    in the case of the  hyper-simple-cubic lattices. }
\begin{tabular}{|l |l |}
 \hline
$c_1$=  1/8& $c_6$=20815/21504 \\
$c_2$= 5/96  & $c_7$=9151/6144\\
$c_3$=  5/64 & $c_8=$ 39593/73728 \\
$c_4$=237/1280 & $c_9=$-645691/61440\\
$c_5$=349/768  &$c_{10}$=-107753037/901120\\
 \hline
\end{tabular} 
\label{tab1}
\end{table}
Viewing the sequence of $c_i$, we are certainly led to expect the sum
in Eq.(2) to be asymptotic and not convergent.

If we consider covering by dimers of a fraction of the vertices
 denoted here by $p=2\rho$ (where $\rho$ is the dimer density per site
 and the vertices not covered by dimers are considered covered by
 monomers(checkers)) and as above study the number of such coverings,
 we arrive similarly at a function $\lambda_d(p)$ where
\begin{equation}
 \lambda_d(1)=\lambda_d
\label{lasp}
\end{equation} 
Another common notation for $\lambda_d$ is $\tilde{h}_d$. One also studies
\begin{equation}
h_d=\max_{0 \le p \le1} \lambda_d(p) .
\end{equation}
For $\lambda_d(p)$ Friedland et al.\cite{6,7} proved the asymptotic relation
(asymptotic as $d \to \infty$) 
\begin{equation}
 \lambda_d(p) \sim \frac{1} {2} ( p{\rm ln}(2d)
-p{\rm ln}(p)-2(1-p){\rm ln}(1-p)-p) .
\label{lasp1}
\end{equation} 
Both this equation and Eq.(1) may be viewed as the mean field
approximations for the respective quantities. This was first mentioned
in Ref.[\onlinecite{14}] and is briefly discussed at the end of Section
III.  By a similar development to that in Ref.[\onlinecite{14}] one of
the authors, P.F. and Friedland \cite{13} argued for an expansion
\begin{equation}
 \lambda_d(p)=\frac{1} {2}(p{\rm ln}(2d)-p{\rm ln}(p)-2(1-p){\rm ln}(1-p)-p)
 +\sum_{k=2}^{\infty} a_k(d)p^k
\label{laspd1}
\end{equation} 
where, setting $x(d)= \frac{1}{2 d}$, those authors computed the
following six coefficients,

$a_2(d)= \frac{1}{4} x$

$a_3(d)= \frac{1}{12} x^{2}$

$a_4(d)= \frac{1}{24} x^{2} \left(- 5 x + 3\right)$

$a_5(d)= \frac{1}{40} x^{3} \left(- 39 x + 20\right)$

$a_6(d)= \frac{1}{60} x^{3} \left(- 19 x^{2} - 30 x + 20\right)$

\noindent The main result of this paper is the extension of known values:

$    a_{7}(d)= \frac{1}{84} x^{4} \left(1093 x^{2} - 1008 x + 231\right)$

$    a_{8}(d)= \frac{1}{112} x^{4} \left(967 x^{3} - 35 x^{2} - 602 x + 189\right)$

$    a_{9}(d)= \frac{1}{144} x^{5} \left(- 66047 x^{3} + 68712 x^{2} - 23556 x + 2856\right)$

$    a_{10}(d)= \frac{1}{180} x^{5} \left(- 67721 x^{4} + 18495 x^{3} + 29565 x^{2} - 15405 x + 2232\right)$

$a_{11}(d)= \frac{1}{220} x^{6} \left(5456221 x^{4} - 6452710 x^{3} + 2752860 x^{2} - 524700 x + 39710\right)$

$a_{12}(d)= \frac{1}{264} x^{6} \left(887437 x^{5} + 2477970 x^{4} - 3847316 x^{3} + 1824724 x^{2} - 378004 x + 31130\right)$

$a_{13}(d)= \frac{1}{312} x^{7} (- 614279535 x^{5} + 794742624 x^{4} -
392705664 x^{3} + 95702984 x^{2}$ \\$
 - 11868441 x + 621504)$

$ a_{14}(d)= \frac{1}{364} x^{7} (678357525 x^{6} - 1192936836 x^{5} + 869146005 x^{4} - 339116960 x^{3} + $\\$
75444460 x^{2} - 9220393 x + 497016)$

$a_{15}(d)= \frac{1}{420} x^{8} (89365899701 x^{6} - 124219633888 x^{5} + 68478916835 x^{4} - 19687487260 x^{3} + $\\$
3185117250 x^{2} - 281248772 x + 10870055)$

$a_{16}(d)= \frac{1}{480} x^{8} (- 206929670185 x^{7} + 330409603725 x^{6} - 221634792330 x^{5} + 83075676915 x^{4} - $\\$
19146441210 x^{3} + 2751382878 x^{2} - 231206020 x + 8907885)$

$a_{17}(d)= \frac{1}{544} x^{9} (- 16388790941183 x^{7} + 24197151077904 x^{6} 
- 14547689415128 x^{5} + $\\$ 
4724677127184 x^{4} - 911997832372 x^{3} + 106422324240 x^{2} 
- 7073226040 x + 210678416)$

    $a_{18}(d)= \frac{1}{612} x^{9} $($55311212276891 x^{8} - 89669360611981 x^{7} 
+ 61471303146642 x^{6} $\\$
 - 23833002227449 x^{5} + 5824219780656 x^{4} - 933123781978 x^{3} 
+ 97025317251 x^{2} - 6063514389 x + $\\$ 
176829104)$

$a_{19}(d)= \frac{1}{684} x^{10} (3770925296332945 x^{8} - 5844092886538362 x^{7} + 3760855236979965 x^{6}$\\$ - 1340101438257204 x^{5}
 + 293876531465913 x^{4} - 41181769780866 x^{3} + 
3649368222699 x^{2}$\\$ - 189574974180 x + 4489042410)$

$a_{20}(d)= \frac{1}{760} x^{10} (- 16045042327489089 x^{9} + 26850617367263509 x^{8} - 19173445082939896 x^{7} + $\\$ 
7825625528101485 x^{6} - 2044727194575071 x^{5} + 359651992720132 x^{4} -
43125672212794 x^{3} $\\$ + 3440152700645 x^{2} - 167626520550 x + 3849436062 )$

For the hsc lattices of dimensions $d < 5$, four more coefficients
$a_{k}(d)$ are available. They are listed in Table \ref{tab2}.
\begin{table}[ht]
\tiny
\caption{ Higher order expansion  coefficients $a_k(d)$ of the dimer entropy 
$\lambda_d(p)$ on the hsc lattices of dimension $d<5$. }
\begin{tabular}{| l| l| l|}
 \hline
 $d=2$ &$d=3$ & $d=4$\\
 \hline
$a_{21}(2)$=255640084561/923589767331840 & $a_{21}(3)$=
66223472491867/1023724363217633280 &$a_{21}(4)$=
15299547547784641/968454063869751459840   \\
$a_{22}(2)$=50273131919/193514046488576 &
 $a_{22}(3)$=1171503630290797/20269742391709138944 &
$a_{22}(4)$=117431629955187175/8522395762053812846592 \\
$a_{23}(2)$=4312434281365/17803292276948992& 
$a_{23}(3)$=6903357438819689/133201164288374341632 &
$a_{23}(4)$= 902716034982108733/74672420010376264941568  \\
$a_{24}(2)$=5789230773063/25895697857380352 & $a_{24}(3)$=
40662370356724697/871862166251177508864 
&$a_{24}(4)$=6949151047607061613/651686574636011039490048 \\
 \hline
\end{tabular} 
\label{tab2}
\end{table}
\normalsize
In Ref.[\onlinecite{13}] it was conjectured that the 
series in Eq.(\ref{laspd1}) is
convergent for $0 \le p \le 1 $.  Author P.F. in fact proved\cite{17}
that this series converges for small enough $p$. 

Using also:

i) the result by Heilmann and Lieb \cite{H-L} that $\lambda_d(p)$ 
is analytic for $0 < p < 1$,

ii) the conjecture that {\it the $a_k(d)$ are all positive for integer values 
of $d$ in the case of the hsc lattices} (that we have checked for integer 
values of $d$ and $k \le 20$, see also the Appendix),

iii) the theorem that, for an analytic function represented in a
vicinity of the origin by a power series with positive coefficients,
one among the singularities nearest to the origin lies on the positive
real axis,
   
we can extend the analyticity domain of  $\sum_k a_k(d) p^k$ to a disk of 
radius $R < 1$. 
The convergence of this series also at $p = 1$ is then a trivial consequence 
of the positivity conjecture for the coefficients $a_k(d)$ and of the upper 
bound \cite{1} $\lambda_d(1) < \frac{\rm{ln}(2d)!}{4 d}$ .

In this paper we assume the validity of the positivity conjecture, from
which the convergence of the series $\sum_k a_k(d) p^k$ for $0 \le p \le 1$
follows.

Since for any $r$, the partial sums $\sum_k^r a_k(d) p^k$ are positive
for integer values of $d$, the expansion Eq.(6) gives good
approximations of $\lambda_d(1)$ also in low dimensions, unlike the
expansion Eq.(2), which is numerically useful only for sufficiently
large $d$.

In the Appendix we shall further discuss the positivity conjecture, 
while Section IV is devoted to the numerical approximations.

It is interesting to
point out some results of historic importance for the dimer problem.
The exact value of $\lambda_2$  calculated by M.E. Fisher\cite{10}
 and P.W. Kasteleyn\cite{11} is given by the closed form expression
\begin{equation}
  \lambda_{2}\equiv \tilde h_2=\frac{1}{\pi}( \frac{1}{1^2}
- \frac{1}{3^2}+ \frac{1}{5^2}- \frac{1}{7^2} \cdots)= G/\pi= 0.2915609040...
\label{lad3}
\end{equation} 
 with $G$  Catalan's constant.
The technique used in the proof of this relation had great influence
 in the field of exactly soluble models.  

The one-dimensional problem
 has an even more complete solution\cite{6}
\begin{equation}
\lambda_{1}(p)=\frac{p} {2}  {\rm ln}(2)-
\frac{p}{2}{\rm ln}(p)-(1-p){\rm ln}(1-p)
-\frac{p}{2}+\sum_{k=2}^{\infty} \frac{(p/2)^k}{(k-1)k}
\label{lad1}
\end{equation}
 so that $\lambda_1(1)\equiv \tilde h_1=0$ and $h_1={\rm ln} \frac
{1+\sqrt 5}{2}$.
Notice that in this simple case, all the $a_k(1) $ are positive and
rapidly vanishing as $k \to \infty$, so that the series converges for $0
\leq p \leq 1$.  

Let us turn for the moment to consideration of a dimer
gas on our $d$ -dimensional lattice.  The gas of dimers is taken as a
``hard body'' system. Between each two dimers there is a potential
energy $0$ if the dimers are disjoint and $+\infty$ if they
overlap. For this gas we are interested into the coefficients of the
Mayer series\cite{ruelle} $ b_1(d), b_2(d),..$.

Both the formalism in Ref.[\onlinecite{14}] used to derive Eq.(2) and
the formalism in Ref.[\onlinecite{13}] used to derive
Eq.(\ref{laspd1}) take as inputs the $b_i(d)$ and have as outputs the
$c_i$ of Table \ref{tab1} and the $a_k(d)$.  Author P.F. did not have
as good an algorithm for computing to high orders the $b_i(d)$ as in
Ref.[\onlinecite{bp1,bp2,bp3}] and was not aware of the already
existing lower-order expansions\cite{gaunt,kurfis,kenzie} for small
lattice dimensions.  This explains the many additional terms
computed in Eq.(2) and Eq.(\ref{laspd1}) when the computations of
Ref.[\onlinecite{bp3}] were used as inputs.  In Sect.II, the technique
used in Ref.[\onlinecite{bp2,bp3}] to compute the $b_i(d)$ is
discussed. For the computations of the $ a_i(d)$, with $ i = 1,2,...20$,
one needed exactly the $b_i(d)$, for $1 \le i \le 20$ and $1\le d \le
10$.  (Interestingly, these values in fact determine, for all $d$, the
$b_i(d)$ with $1 \le i \le 20$.  This will be shown in Section II and
in an independent way in Section III.)

In Sect.III the machines in Refs.[\onlinecite{14}]  and
[\onlinecite{13}] to calculate the $c_i$ and $a_k(d)$ respectively, are
discussed. But they are too technical to get deeply into all of the
theory. Recently, in fact within the past year, P.F. found another
route from the $b_i(d)$ to expansions for $\lambda_{d}(p)$, simple
enough for us to completely describe it in this paper\cite{12}.
We close this Section by specializing\cite{15} the expansion in
Eq.(\ref{laspd1}) to $d=2$, to see what such an expansion looks like 
\begin{equation}
 \lambda_2(p) = \frac{1} {2}(p{\rm ln}(4)-p{\rm ln}(p)-2(1-p){\rm ln}(1-p)-p)
+ 2(\frac{1}{2\cdot 1}(\frac {p} {4})^2+\frac{1}{3\cdot 2}(\frac {p} {4})^3+
\frac{7}{4\cdot 3}(\frac {p} {4})^4+ \frac{41}{5\cdot 4}(\frac {p} {4})^5\cdots
\label{lad2}
\end{equation} 
where this equation is determined by an infinite sequence of integers
\begin{equation}
1,1,7,41,181,757,3291,14689,64771,276101,1132693,4490513,17337685,...
\label{square}
\end{equation}
of which the first 23 integers are known from the calculations of this
paper.  It is very natural to try to find a pattern in the successive
terms of this sequence so that a closed form expression for
$\lambda_2(p)$ be realized, recalling that it exists for $\lambda_2
\equiv \lambda_2(1)$. 

Recently, we came across an early paper by Rushbrooke, Scoins and
Wakefield [\onlinecite{rush}] computing by a somewhat different method
the first six coefficients in Eq.(\ref{laspd1}) for the square 
and the diamond lattices and the first five for other  three-dimensional
lattices.

The rest of the paper is organized as follows.  In Section II we
recall how the Mayer expansion for the dimer problem is related to the
high-temperature(HT)  low-field expansion of an Ising system.  Section III
sketches how the expansion of Eq.(\ref{laspd1}) is derived from the
Mayer series.  In Section IV we show how simply the expansion
Eq.(\ref{laspd1}) can lead to accurate estimates of the $p$-dependent
entropy $\lambda_d(p)$.  The Appendix contains additional comments on
the positivity conjecture and lists the coefficients appearing in some
generalizations of Eq.(\ref{laspd1}) to lattices other than the
hsc. We have included in the Appendix  a last subsection on the
graphical expansion procedure for the Ising model, that completes the
exposition of Section II.

\section{Dimers and the Ising model}
It has long been known\cite{10,11,gaunt,kurfis} that the number of
ways to place $s$ hard dimers onto a lattice can be evaluated 
 by computing, to the same order $s$ and on the same lattice,
the HT and low-field series expansion of the
free-energy of a spin-$1/2$ Ising model in the presence of a uniform
magnetic field.

The dimer combinatorial problem can be simply formulated in the
language of statistical mechanics. A set of dimers on a $N$-site lattice
($N$ even) is described as a lattice-gas of molecules occupying
nearest-neighbor sites, subject to a non-overlap constraint, in terms
of a macro-canonical partition function
\begin{equation} \Xi_N(z)=1+
\sum_{s=1}^{N/2} Z_sz^s= 1+
\sum_{s=1}^{N/2} g_N(s)z^s.
 \label{mdmpfNZ}
\end{equation}
Due to the non-overlap constraint, $Z_s$, the canonical partition
function for a fixed number $s$ of dimers, simply counts the allowed
dimer configurations, so that $g_N(s)$ is precisely the number of ways
of placing $s$ dimers over the links of the lattice, and
$z=\exp{(\beta\mu)}$ is the dimer activity. The chemical potential $\mu$,
namely the energy cost of adding one more dimer to the system, is zero
whenever there is room on the lattice for adding one more dimer and
infinite otherwise.  Therefore the value of $\beta=1/k_BT$ with $T$
the temperature and $k_B$ the Boltzmann constant, is irrelevant and
can be fixed to unity. Thus $z=1$ is the value of the activity
describing the combinatorics of a monomer-dimer system i.e. of a dimer
system that does not cover completely the lattice, while $z=\infty$
describes the complete coverings.

  In the
$N \to \infty$ (thermodynamical) limit one gets
\begin{equation}
\Xi(z)=\lim_{N\rightarrow \infty}[\Xi_N(z)]^{1/N}
=1+  \sum_{s=1}^{\infty} g(s)z^s
\label{mdmpf}
\end{equation}
from which a ``pressure" (or macro-canonical potential) can be defined
in the usual way
\begin{equation}
 P(z)={\rm ln}(\Xi(z))
= \sum_{s=1}^{\infty}b_sz^s
\label{mpot}
\end{equation}
since $\beta=1$.
The dimer density per site $\rho$ is expressed in terms of the pressure by
\begin{equation}
\rho(z)=z \frac{dP} {dz}
= \sum_{s=1}^{\infty}sb_sz^s.
\label{mden}
\end{equation}
The series for $\rho(z)$ can be inverted to get $z$ as a power series
in the density and by substituting $z=z(\rho)$ in Eq. (\ref{mpot}),
$P$ can be expressed as a power series in the density $\rho$, thus
obtaining the virial expansion.  Eqs.(\ref{mpot}) and (\ref{mden}) are
called the Mayer expansions of the dimer lattice-gas.

The specific entropy $s_d(p)$ of a dimer system of density $\rho$ in
$d$ dimensions is
\begin{equation}
s_d(p)/k_B \equiv \lambda_d(p)=-\rho(z) {\rm ln} z +P(z)=
\frac{1} {2}(p{\rm ln}(2d)-p{\rm ln}p) +O(p)
\label{mentr}
\end{equation}
where the last expression arises by setting $z=z(p)$, and $\rho=p/2$, and observing
that on the hsc lattices $z=\frac{p}{2d}+O(p^2)$.

Notice that one has

\begin{equation}
    \frac{d \lambda_d}{d p} = \frac{-{\rm ln}(z)}{2}.
    \label{dlambda}
\end{equation}

This structure was further specified in Ref.[\onlinecite{6,7}], as indicated in
Eq.(\ref{laspd1}).  
 One can also easily check that changing the variable from $z$ to $p$,
the point $z=1$ corresponds to a stationary point of the entropy with
respect to $p$, thus linking the definition given above of $h_d$ in
terms of $\lambda_d(p)$ with the definition used in
Ref.[\onlinecite{bp1}] as $P(z)|_{z=1}$.
We now couple the relation Eq.(\ref{mentr}) with the expansions above
for $P(p)$ and $z(p)$. We write 
\begin{equation}
z=\frac{p}{2b_1} (1+F(p))
\label{3.6}
\end{equation}
and then get from Eq. (\ref{mentr}) and (\ref{3.6})
\begin{equation}
\lambda_d(p)=P(p)-\frac{p}{2}{\rm ln}(\frac{p}{2b_1})
-\frac{p}{2}{\rm ln}
 (1+F(p))
\label{3.7}
\end{equation}
or 
\begin{equation}
\lambda_d(p)=P(p)-\frac{p}{2}{\rm ln}(p)+ \frac{p}{2}{\rm ln}(2d) -\frac{p}{2}{\rm ln}(1+F(p))
\label{3.8}
\end{equation}
 using $b_1=d$. 
 Referring to Eq.({\ref{laspd1}}) we may put Eq. (\ref{3.8}) in the form
\begin{equation}
\lambda_d(p)=\frac{1}{2}(p{\rm ln}(2d)-p{\rm ln}p -2(1-p){\rm ln}(1-p)-p)+
\sum_{k=2}^{\infty}a_k(d)p^k
\label{3.9}
\end{equation}
where the $a_k(d)$ are suitably determined from the Mayer series
coefficients in a straightforward manner.

The Mayer coefficients for the dimer system $b_s(d)$ on a
$d$-dimensional lattice are simply obtained from the HT expansion of
the free-energy for the Ising model.  To illustrate the relationship
between the Ising and the dimer problems, recall the
``primitive''\cite{domb} method of HT and low-field graphical
expansion for the partition function $Z_N(\beta,h)=\sum_{m \ge
0}\sum_{l=m}^{L_{max}}\gamma_N(2m,l){\rm tanh}(h)^{2m}{\rm
tanh}(\beta)^l $ of a spin-1/2 Ising model on a lattice of $N$
sites. Here $\beta=1/k_BT$ denotes the inverse temperature and
$h=\beta H$ with $H$ the uniform external magnetic field.  The
expansion coefficient $\gamma_N(2s,s)$ counts all possible lattice
configurations of graphs represented by precisely $s$ disconnected
edges placed onto disjoint links of the lattice and therefore
coincides with the quantity $g_N(s)$ in Eq.(\ref{mdmpf}). The
procedure of forming the specific free-energy
$f_N(\beta,h)=\frac{1}{N}{\rm ln} Z_N$, and then taking the
thermodynamical limit, exactly parallels\cite{kurfis} the procedure
leading to Eq.(\ref{mpot}), so that one concludes that from the
expansion $f(\beta,h)=\sum_{m \ge 0}\sum_{l=m}^{L_{max}}f_{2m,l}{\rm
tanh}(h)^{2m} {\rm tanh}(\beta)^l$, the Mayer expansion coefficients
can be read as $b_s(d)=f_{2s,s}(d)$.

Let us now recall that recently a significant extension, of the HT
series for several models in the Ising universality class, including
the conventional spin-$1/2$ model, has been obtained for a sequence of
bipartite lattices, in particular the hsc lattices of spatial
dimension $1\le d \le 10$ and the hyper-body-centered-cubic
(hbcc) lattices of any dimension. In the case of the hbcc lattice, 
this is true at least in principle, because the lattice dimension
enters only in the power of the embedding number (see below), and thus
the computation time increases very slowly with the dimension; so
far we have only performed the computations for $d \leq 7$.  It is
also convenient, at this point, to give some simple details on these
calculations. It is most convenient to refer\cite{bp2,bp3} to the
classical linked-cluster method\cite{wortis} of graphical expansion.
At each order $l$ of HT expansion, the series coefficients are
expressed as the sum of an appropriate class of $l$-edge graphs. Each
graph contributes a ratio of two integers: the
``free-embedding-number'' and the symmetry-number of the graph, times
a product of ``bare vertex-functions'' associated to the vertices of
the graph and depending on the magnetic field. The embedding-number
counts the number of distinct ways (per site of the underlying
lattice) in which the graph can be placed onto the lattice, with each
vertex assigned to a site and each edge to a link. This number depends
on the topology of the graph and on the dimension $d$ of the
lattice. The important property is that, in the case of the hsc
lattices (but not for the hbcc lattices!), for a generic graph with
$l$ edges, the embedding-number is a polynomial in $d$ of degree $l$
at most.  The symmetry-number counts the automorphisms of the graph
and depends only on the topology of the graph.  The great advantage of
the linked-cluster method comes from the recognition that the huge
variety of graphs that contribute at relatively high orders of
expansion to the computation of a physical quantity, e.g. the
magnetization, can be obtained by combining simpler graphs in a
smaller class\cite{wortis}, thus making possible to trade the
computational complexity for algebraic complexity.

From the field-dependent free-energy, one can compute all its
field-derivatives usually called (higher) susceptibilities. It is
clear at this point that, on the hsc lattices, the computation of
these quantities through the 10th order, can be extended to a generic
$d$. It is sufficient to perform a simple interpolation of the series
coefficients using the computation on a sequence of hsc lattices of
dimensions $1 \le d \le 10$ and basing on the fact that the
$l$th order expansion coefficient is a simple polynomial\cite{fisga}
of degree $l$ in $d$ (with zero constant term).  Actually much more
than this can be done. One can observe\cite{bp3} that the knowledge of
the free-energy gives access to the HT expansions of the successive
derivatives of the magnetic field with respect to the magnetization
$\partial^{2p+1} h/\partial {\cal M}^{2p+1}$, for $p=0,1...$ and that
these quantities are expressed only in terms of connected graphs
having no articulation-vertex, i.e. no vertex whose deletion would
disconnect the graph.  What is decisive for our aims is the fact that
the embedding-number onto a hsc lattice of an $l$-edge graph in this
particular class is a polynomial in $d$ of degree $\lfloor l/2\rfloor$
at most\cite{fisga}. Here $\lfloor l/2\rfloor$ denotes the integer
part of $l/2$. Therefore, in spite of the fact that the HT expansion
coefficients of the (higher)-susceptibilities at order $l$ are
polynomials in $d$ of degree $l$, the susceptibilities can be simply
expressed in terms of the successive derivatives of the magnetic field
with respect to the magnetization which, at the same expansion order,
are polynomials in $d$ of degree $\lfloor l/2\rfloor$ only.  Thus, one
can conclude that the exact dependence on $d$ of the HT coefficients
of the higher susceptibilities can actually be determined up to order
20, using only data for a sequence of hsc lattices of dimensions
$1 \le d \le 10$, by an interpolation in $d$ of the series
coefficients.

Let us finally stress that the elements of the coefficients matrix
$f_{2m,l}(d)$ of the HT and low-field expansion for the free-energy of
the spin-1/2 Ising model can be linearly expressed in terms of the
expansion coefficients of the susceptibilities and therefore they also
are polynomials in $d$ of degree $l$.  This property holds in
particular for the Mayer coefficients $b_s(d)=f_{2s,s}(d) $ of the
dimer gas. 
From Eq.(\ref{dlambda}) it follows that $\frac{d \lambda_d}{d p}$ can also be 
determined to the order 20 for all $d$.

More details concerning the graphical expansion procedure
 can be found in Subsect. C of the Appendix.
\section{Derivation of Expansions}
As mentioned in the introduction, we have a second route
 for deriving $\lambda_d(p)$ and
$\lambda_d$ expansions. The key initial step is the computation of the
quantity $\tilde J_i(d)$ from the quantities $b_i(d)$.  The $\tilde
J_i(d)$ depend on the set of $b_n(d)$ with $n \le i$.  The
computations are given in Ref. [\onlinecite{20}] as follows
\begin{equation}
\nonumber
\tilde J_1=0.
\end{equation}
We first find $\tilde J_r^L$, with $\tilde J_1^L=0$, and from $r=2$
on, inductively defined by
\begin{equation}
\tilde J_r^L=\frac{1}{L}\Big\{ S_r-(exp(L\sum_{i=1}^{r-1}\tilde J_i^Lx^i))|_r\Big\}
\label{3.10}
\end{equation}
where
\begin{equation}
S_r=\sum_{p=0}^r\Big\{ (exp(L\sum_ib_i(\frac{x}{2d})^i))|_p\frac{1}{(r-p)!}
\big (\frac{-1}{2(L-1)} \big )^{r-p} \frac{(L-2p)!}{(L-2r)!}\Big\}.
\label{3.11}
\end{equation}
The symbol $|$ with the subscript $j$ indicates the $j$th coefficient
 in the formal power series in $x$. The $\tilde J_r$ are determined from
 the $\tilde J_r^L$ by taking $L$ to infinity.  We may also
 inductively go from the $\tilde J_i$ to the $b_i$ by the same
 formulae.

 This set of relations was first implicitly used in
Ref.[\onlinecite{14}], but not explicitly written down there. Just as
the $b_i(d)$ are the cluster expansion coefficients of a dimer gas,
the $\tilde J_i(d)$ are the cluster expansion coefficients of a
certain polymer gas\cite{14} and these coefficients of the two gases
are related by the development surrounding eqs. (\ref{3.10}) and
(\ref{3.11}).  This is a clean calculation that requires no hard
proof. The $\tilde J_i(d)$ can be proved \cite{3} to be of the form
\begin{equation}
\tilde J_s(d)=\frac{c_{s,r}}{d^r}+\frac{c_{s,r+1}}{d^{r+1}}+...+\frac{c_{s,s-1}}{d^{s-1}} 
\label{3.12}
\end{equation}
with $r \ge s/2$.

Whereas our first development was basically for each $d$ individually,
we will see as with this last equation that the dependence on $d$ is in
the nitty-gritty of this second development. The present treatment
allows us to get results relating the series for different $d$'s.  As
an example, suppose we know the $\tilde J_i(d)$ for $1 \le d \le 10$,
$i \le 20$.  Then one may derive $\tilde J_i(d)$ for $i \le 20$ and
all $d$! (One has enough information to compute all the $c_{s,r}$ for
$i \le 20$.) The same statement holds for the $b_i(d)$, since one may
go between the set of $b_i(d)$ with $i < n$ and the set of $\tilde J_i(d)$
with $i<n$, as mentioned above.

So far all the results dealt with in this section have been true and
rigorously proven. We now turn to the further development, certainly
true, but for which we do not yet have a rigorous proof.  We work for a
given $d$ and take as known the $\tilde J_i(d)$ (which as above could
be calculated from the $b_i(d)$). We then compute $\alpha_i(d)$ by
iterations, from $\alpha_i=0$, of 
\begin{equation}
\alpha_k= \big (\tilde J_kp^k\big ) \frac{1}{(1-2\sum_{i=2} i\alpha_i)^{2k}}\big( 1-2\sum_{i=2} i\alpha_i/p \big)^k. 
\label{3.13}
\end{equation}
In iterating, we take the mapping from the right side of the equation to
the left side of the equation to be a mapping of formal power series in $p$.
It is proven in Ref.[\onlinecite{17}]     that there is an $m > 0$ such that each of the sequences of
formal power series converges to a convergent power series of radius
of convergence $ \ge{m}$. (Even if the power series in $\lambda_d(p)$, see Eq.(\ref{laspd1}),
has a radius of convergence $\ge{1}$, as we assume, we do not know if $m$ can be picked
to be 1.) Then $\lambda_d(p)$ is given by 
\begin{equation}
\lambda_d(p)=Q_1+Q_2
\label{3.14}
\end{equation}
\begin{equation}
Q_1=\frac{1}{2}(p{\rm ln}(2d)-p{\rm ln}p -2(1-p){\rm ln}(1-p)-p)
\label{3.15}
\end{equation}
\begin{equation}
Q_2=\sum_{i=2} \alpha_i-\sum_{k=2}\frac{1}{k}\Big ( 2\sum_{i=2}i
\alpha_i\Big)^k+
\frac{1}{2} p\sum_{k=2}\frac{1}{k}\Big ( 2\sum_{i=2}i\alpha_i/p\Big)^{k}.
\label{3.16}
\end{equation}
$Q_2$ may be developed as a power series in $p$ 
\begin{equation}
Q_2=\sum_2^{\infty} a_k (d) p^k
\label{3.17}
\end{equation}
where $a_k(d)$ is a polynomial in powers of $\frac{1}{d}$ with powers
satisfying $k/2 \le r < k$ (as the powers in Eq. (\ref{3.12})), see
Refs.  [\onlinecite{13}] or [\onlinecite{17}].  So for example if we
know $a_k(d)$ for $k \le 20$ and $d \le 10$, then we can deduce
$a_k(d)$ for  $k \le 20$ and all $d$.
To determine $a_k$ only the values 
of $a_k(d)$,  $d=1,..,\lfloor k/2 \rfloor$ are needed,
 the remaining $10 - \lfloor k/2 \rfloor$ values were used to give consistency
 checks for each  $k < 20$. 
This is a consistency check on both the computation 
 of the $b_i(d)$ and of the theory, since as we mentioned above the 
 development of Eq.(\ref{3.13})-Eq.(\ref{3.17}) has not been yet made
 rigorous.

We can deduce the series for $\lambda_d$, Eq. (2) above, basically by
 setting $p=1$ in Eq.(\ref{3.14}). It is important to note for this
 that each power of $\frac{1}{d}$ gets a contribution from only a
 finite number of $a_k(d)$. Specifically $1/d^s$ get contributions
 from those $a_k(d)$ for which $k/2 \le s < k$. For example if we know
 $a_k(d)$ for $k \le 20$, then we can deduce the terms in $\lambda_d$
 up to $1/d^{10}$.
 
 To get at the theory (of the formal argument leading to
 eqs.(\ref{3.14})-(\ref{3.17}), our second development of the
 $\lambda_d(p)$ and $\lambda_d$ expansions), we recommend to the
 reader starting by reading Ref.[\onlinecite{14}], a three page paper,
 or Section 5 of Ref.[\onlinecite{13}].  We now give a slightly hand
 waving capsule summary of the introductory portion of this theory up
 to the derivation of the mean field formulae Eq.(1) and
 (\ref{lasp1}) above.
 
We work on a periodic $d$-dimensional lattice with number of sites  $N$. A
``difunction'' is a translation invariant periodic function on pairs
of distinct vertices.  We associate to dimers the difunction $f$, that
is $1$ if the pair of vertices are nearest neighbors and $0$
otherwise.  We call a sequence $ X_1, X_2,....$ of $pN$ distinct
vertices a ``$p$-sequence''. We let $\sum$ denote the sum over all 
$p$-sequences. We note that the number of distinct dimer coverings that 
cover a fraction $p$ of the vertices can be represented as 
\begin{equation}
\frac{1}{2^{(pN/2)}} \frac{1}{(pN/2)!} 
\sum \mathop{\prod_{i=1}^{pN}}_{i\text{ odd}} f( X_i, X_{i+1}).
\label{3.18}
\end{equation}
The numerical factors before the sum divide by the number of different
$p$-sequences that correspond to the same choice of dimers.
The sum is over $\frac{N!}{((1-p)N)!}$ $p$-sequences.

We let $f_0$ be the difunction of constant value $(\frac {2d}{N-1})$. $f$
and $f_0$ have the same ``normalization'' in the sense that, if one
fixes its first component and sums over the second, one gets the same
answer for both functions.  Replacing $f$ in Eq.(\ref{3.18}) by $f_0$
and using the Stirling formula gives the mean field answer
\begin{equation}
\nonumber
\exp \lambda_{mf}N
\end{equation}
for the number of our dimer covers, where $ \lambda_{mf}$ is as in
Eq. (\ref{lasp1}).

We write
\begin{equation}
f=f_0+ \cal V
\label{3.19}
\end{equation}
with 
\begin{equation}
{\cal V} =f-f_0.
\label{3.20}
\end{equation}
Expansions in powers of $ \cal V$ may be converted into the expansions 
of this paper.
\section{Numerical estimates}
It is interesting at this point to get some feeling about the accuracy
of the estimates of $h_d$ and $\tilde h_d \equiv \lambda_d(1)$ that
can be obtained from the expression Eq.(\ref{laspd1}) for
$\lambda_d(p)$ when a sufficiently large number of coefficients
$a_k(d)$ is known.  For the evaluation of both $h_d$ and $\tilde h_d $
a first orientation comes from truncating the expansion
$\sum_{k=2}^{\infty}a_k(d)p^k$ at the order $k=r$ and plotting the result
vs. some power of $1/r$.  Let us first consider the quantity $\tilde
h_d$. Assume that the series converges for $p=1$ and that the
coefficients $a_k(d)$ are all positive, then its successive
truncations must provide an increasing sequence of lower bounds of the
limit. First, we can check that the approximation of truncating the
expansion at the highest known order is always consistent with the
known upper bounds. However in the case of $d=2$ and $d=3$ the values
thus obtained, i.e. $\tilde h_2=0.2865..$ and $\tilde h_3=0.44916..$
respectively, appear to be still too small.  Therefore one should
properly extrapolate the sequences $(S_r)=\frac{1} {2}({\rm ln}(2d)-1)
+ \sum_2^{r}a_k(d)$ of the truncated expressions. Of course, the
best way to do this depends on the behavior of the sequences. It is
very encouraging to notice that for all values of $d$, the sequences
are smooth and their behavior is well approximated by the simple Ansatz
 $S_r= a + b/r^{\alpha}$ and so one has $a \approx \tilde h_d$. This
procedure is very successful. We observe that  $\alpha$   increases with
$d$ and  ranges from $\alpha \approx 1$ for $d=2$ to $\alpha \approx
2.6$ for $d=6$. In dimension $d=2$, this Ansatz gives a good fit
of the last 4 - 10 terms of the sequence and the extrapolated value
$a=0.2915(20)$ agrees with the exactly known value $\tilde
h_2=0.291560...$ in Eq.(\ref{lad3}), within the estimated error. The
uncertainty we have written is very conservative, although somewhat
arbitrary. It is obtained both allowing for the spread of values resulting
from small variations of the exponent $\alpha$ in the functional form
used for fitting and from a comparison with other extrapolations 
obtained for example, evaluating the series $\sum_{k=2}^r a_k(d)p^k$
for $p=1$, by Pad\'e or differential approximants\cite{gutt} and
adding the result to the expression $\frac{1} {2}({\rm ln}(2d)-1)$.
Analogously, for $d=3$ the sequence $(S_r)$ is well fitted by the
Ansatz $a+b/n^{1.3}$ and leads to the estimate $\tilde
h_3=0.4499(2)$. This value is not far from the estimate $\tilde
h_3=0.4479$ obtained by a MonteCarlo calculation\cite{beichl} or from
$\tilde h_3=0.453(1)$ obtained\cite{21} extrapolating a much shorter
expansion, and it is also completely consistent with the known 
bounds\cite{1,6,schrij,lund} $0.440075842 \leq \tilde h_3 \leq 0.4575469308$.

Proceeding along the same lines, we
 can determine the values of $\tilde h_d$ for any value of $d$. We
 notice that the apparent precision of the results improves rapidly
 as $d$ grows, while the differences between the extrapolated values
 and the highest order truncations of the series (as well as the estimated
 uncertainties)  decrease rapidly. The final estimates are always completely 
 consistent with the known bounds. Our estimates of $\lambda_d(1)$
 for $d=2,...,8$ are reported in table \ref{tab3}. Notice that evaluations
of these quantities appear  rarely\cite{finch} in the literature.

The computation of $h_d$ requires only a quite short comment.
Unsurprisingly, the sequences of truncated expansions $\sum_{k=2}^r
a_k(d)p^k$ evaluated for $p< 1$ show a faster convergence than for
$p=1$.  The estimates of $h_d$ thus obtained agree well, within their
uncertainties, with those already listed in Table VII of
Ref.[\onlinecite{bp1}] which have been obtained resumming via Pad\'e
approximants the expansion of $P(z)$ for $z=1$. Therefore the reader
is referred to this source.
\begin{table}[ht]
  \caption{ Our estimates of $\tilde h_d=\lambda_d(1)$ for the
    (hyper)-simple-cubic lattices of dimensions $d=2,3,...8$ with the
    known rigorous lower and upper bounds\cite{13} defined by
    $(1/2){\rm ln}(2d)-1/2 \leq \lambda_d(1) \leq {\rm ln}((2d)!)/4d
    $.  While these rigorous bounds are valid for all $d$, for $d=2$
    we have simply reported the first eight digits of the exact value
    and for $d=3$ we have reported the tighter bounds from
    Refs.[\onlinecite{1,6,schrij,lund}].  The non-rigorous lower
    bounds are simply obtained assuming the validity of the positivity
    conjecture for the coefficients $a_k(d)$ and truncating our
    expansions at the highest available order.}
\begin{tabular}{|c|c|c|c|c|}
 \hline
 & Lower Bound& Non-rig. L.B.& Our Estimate& Upper Bound\\
 \hline
$\lambda_2(1)$&0.29156090&0.286521& 0.2915(20) &0.29156090\\
$\lambda_3(1)$&0.44007584&0.449164& 0.4499(2)  &0.45754694\\
$\lambda_4(1)$&0.53972077&0.576517& 0.57666(3) &0.66278769\\
$\lambda_5(1)$&0.65129254&0.679434& 0.67949(2) &0.75522063\\
$\lambda_6(1)$&0.74245332&0.765301& 0.765315(2)&0.83280061\\
$\lambda_7(1)$&0.81952866&0.838785& 0.838789(1)&0.89968648\\
$\lambda_8(1)$&0.88629436&0.902947& 0.902949(1)&0.95849563\\
 \hline
\end{tabular}
\label{tab3}
\end{table}
\subsection{Series expansion for $h_d$ for $d$ large}
As $d$ goes to infinity, $h_d$ tends to $\tilde{h}_d$.
One can compute the rate with which the former approaches the latter by 
performing an expansion in $\frac{1}{\sqrt{d}}$.

To compute $h_d$ one looks for a stationary point of Eq.(\ref{laspd1}).
Putting $y = 1/\sqrt{2d}$, the stationarity equation can be written as
\begin{equation}
    (1 - p_{st})^2 - p_{st}y^2 \exp(-2\sum_{k=2}ka_kp_{st}^{k-1}) = 0
\end{equation}
This equation can be solved for large $d$.
Knowing $a_k$ up to $k=20$, one can solve iteratively the equation
up to the order $y^{42}$. Here we shall report only the first few terms
\begin{equation}
p_{st} = 1 - y + \frac{1}{2}y^2 + \frac{3}{8}y^3 - y^4 + \frac{201}{128}y^5 
- \frac{ 5}{2}y^6 + \frac{ 7003}{1024}y^7 - 22y^8 + ...
\label{py}
\end{equation}
At the second order in $y$, it agrees with the value of $p_{st}$ associated to
the lower bound for $h_d$ found in \cite{6,7}
\begin{equation}
  p_{st} = \frac{4d + 1 - \sqrt{8d + 1}}{4d}
\end{equation}
Substituting  Eq.(\ref{py}) into Eq.(\ref{laspd1}) to get $h_d$, and $p=1$ into
Eq.(\ref{laspd1}) to get $\tilde{h}_d$, one finds

\begin{equation}
    h_d - \tilde{h}_d = y - \frac{1}{4}y^2 - \frac{11}{24}y^3 + ...
\label{diffh}
\end{equation}
\begin{equation}
    h_d = \frac{1}{2}(\ln{2d} - 1) + \frac{1}{\sqrt{2d}} - \frac{11}{48\sqrt{2d^3}} + O(d^{-2})
\label{hdas}
\end{equation}
in which we wrote only the first three terms out of the $40$ terms we computed.
Using $40$ terms, this series expansion agrees with the difference
$h_d - \tilde{h}_d$ computed numerically
up to $2\cdot10^{-6}$ for $d=7$, $10^{-15}$ for $d=20$.
From $d=40$ up to $d=9000$ the precision is only $10^{-16}$.

The terms given in Eq.(\ref{hdas}) give $h_d$ with an error less than 
$3\cdot 10^{-3}$ for $7 \leq d < 100$ and 
$2\cdot10^{-5}$ for $100 \leq d < 10000$.

In particular from Eq.(\ref{py}) and Eq.(\ref{diffh}) one gets
\begin{equation}
    \lim_{d \to \infty} \frac{h_d - \tilde{h}_d}{p_{st}(d) - 1} = -1
\end{equation}

\section{Appendix}
\subsection{The conjecture that the coefficients $a_k$ are positive
 in the case of the hsc lattices}
We proved that the coefficients $a_k(d)$ are positive integers for $k
\leq 20$ and $d \geq 1$ by locating in the complex $d$-plane
 their real roots, and counting the
complex ones to make sure that none is missing.  It is interesting to
note that for $1 < d < 2$ or for $2 < d < 3$ the $a_k(d)$ can be
negative, and that there are roots approaching $1$ and $2$ as $k$ gets large.
\begin{table}[ht]
\caption{ Real roots of $a_k(d)$ for $k \geq 10$}
\begin{tabular}{|c|c|c|c|}
 \hline
 $k=10$&  -0.65502486055142554 & 0.99997855862379883 & \\
 $k=11$&  1.0000010707811947 & 1.6603775954637132 & \\
 $k=12$&  -0.12473442164710268 & 1.0000008060184913 & 1.5835444714309055 \\
 $k=13$&  0.99999998817575145 & & \\
 $k=14$&  1.0000000363954472 &1.628126162558255 & \\
 $k=15$&  1.000000000099149 &1.9594209128425236 & \\
 $k=16$&  0.99999999993591589 & & \\
 $k=17$&  0.99999999999934955 & 2.0071302031011769 & 2.4952449117198663 \\
 $k=18$&  1.0000000000002549 & 1.9889667170409254 & \\
 $k=19$&  1.0000000000000033  &1.9993853767904753 & \\
 $k=20$&  0.99999999999999901 & 2.0009898597900763&2.6230186617839066  \\
 \hline
\end{tabular}
\label{tabr}
\end{table}
As we have already noticed, based on the conjecture that the $a_k(d)$
 are positive, the computed values of $a_k(d)$ provide a lower bound of
 $\lambda_d$.  For $d=2$ in the case of $h_d$, this lower bound
 $0.662798966$, is smaller than the estimate $0.662798972(1)$ obtained
 in Ref.[\onlinecite{bp1}] by Pad\'e approximants. For $d \geq 3$
 these lower bounds reproduce within the error the Pad\'e estimates of
 Ref.[\onlinecite{bp1}].
\subsection{Generalization of the positivity conjecture to other
 bipartite lattices}
There is some evidence that the positivity conjecture can be extended
to other {\it bipartite} lattices.  Let us recall what is known on other
lattices.  $\lambda_d$ has been computed from the Mayer coefficients
$b_n$ on other lattices using the formula
\begin{equation}
\lambda_d = -\frac {1} {2} p {\rm ln}(\frac {p} {q}) - (1-p){\rm ln}(1-p)
 - \frac{p}{2} + \frac{q}{2} \sum_{k=2} 
\frac{C_k(\frac{p}{q})^k} {k(k-1)}
\label{other}
\end{equation}
with $q$  the lattice coordination number. The notation $a_k = \frac{q}{2}\frac
{C_k/q^k} {k(k-1)}$ extends that used for the hyper-cubic
case.

From Eq.(\ref{dlambda}) and Eq.(\ref{other}) it follows that
\begin{equation}
    z = \frac{p}{q} exp(-\sum_{k \geq 1} \frac{C_{k+1} - 2 q^k}{k}
    (\frac{p}{q})^k)
\end{equation}
corresponding to Eqs.(9, 21) in Ref.[\onlinecite{rush}], in
which the first few coefficients for the square lattice and for 
some of the lattices discussed below were computed.

Let us now  report the available data for other bipartite lattices.

In the case of the tetrahedral lattice ($q=4$), taking the Mayer coefficients 
$b_n$ from
Ref.[\onlinecite{kurfis,kenzie}] 
we obtain the following set of coefficients $C_k$

$1,1,1,1,31,253,1261,4897,16201,49501,161239,643969,3006823,14104861,60942421,$
$237903169,854124745,2955594097$

In the case of the hbcc lattices, the $b_n$ for $n \leq 24$ have been
computed in Ref.[\onlinecite{bp1}] for $d=3,4,5,6,7$. The coefficients
$C_k$ computed from them are all positive. In Table \ref{bcc}, we list
the coefficients $C_k(d)$ for hbcc lattices of dimensions
$d=3,4,5$.  The coordination-numbers of these lattice are $q=2^d$.
\begin{table}[ht]
\caption{The coefficients $C_k(d)$ in Eq.(\ref{other}) with $ k=2,...,24$ for the 
(hyper)-body-centered-cubic lattices of dimensions  $d=3,4,5$.}
\begin{tabular}{|c|c|c|c|}
\hline
&$d=3$&$d=4$&$d=5$\\
\hline
$C_2$&1 & 1 & 1\\
$C_3$&1 & 1 & 1\\
$C_4$&37 & 151 &541\\
$C_5$&241 & 1001 & 3601\\
$C_6$&1651 & 21241 & 290851\\
$C_7$&13861 & 276445 & 4136581\\
$C_8$&109873 & 4138275 & 185447641\\
$C_9$&850465 & 61222177 & 3766174561\\
$C_{10}$&6620401 & 903139171 & 134478272521\\
$C_{11}$&51657541 & 13527055301 & 3251891481301\\
$C_{12}$&403327651 & 201952069177 & 105463232417731\\
$C_{13}$&3151118881 & 3041256137921 & 2794164743354401\\
$C_{14}$&24647038963 & 45839858214697 & 86840903677417891\\
$C_{15}$&192950685061 & 69396577375846 & 2421252466929163141\\
$C_{16}$&1510882839217 & 10530703348244851 & 73870429278903327001\\
$C_{17}$&11833222518145 & 160247978490447425 & 2123026721471921771521\\
$C_{18}$&92728596423613 & 2444106838568935375 & 64306694719829414761621\\
$C_{19}$&727194198560401 & 37359234126615235321 & 1883895461127373802533921\\
$C_{20}$&5707071682914097 & 572176086489368008851 & 56961277210166888567226841\\
$C_{21}$&44820667959807601 & 8779078842662089743601 & 1690242630478526669835704401\\
$C_{22}$&352227866459521537 & 134925544759538198882283 & 51146624643545703193238849401\\
$C_{23}$&2769671081569110445 & 2076868645293925124133493 & 1531526780518608097927545101821\\
$C_{24}$&21790699297032926587 & 32014374542692855556562921 & 46435767644223061358549293433371\\
 \hline
 \end{tabular}
 \label{bcc}
 \end{table}

For the hexagonal lattice \cite{Ftri} with $q=3$, the coefficients $C_k$ are

$1, 1, 1, 1, 11, 85$

Let us now turn to the case of non-bipartite lattices.

For the triangular lattice ($q=6$) the coefficients up to $C_6$ are listed
in Ref.[\onlinecite{Ftri}], while higher-order ones are obtained from
Ref.[\onlinecite{kurfis}]

$1,-3, -11, 1, 91, 141,-1651,-16143,-87329,-295063,-72533,8092033,76819835$

For the fcc lattice ($q=12$), from Ref.[\onlinecite{kurfis}] we obtain:

$1,-7,19,41,-779,3557,46327,118529,-557909$

These data imply that the positivity conjecture for  the $C_k$
 has to be restricted to bipartite lattices.

 On a Bethe lattice\cite{nagle,stilck} the entropy is given by Eq.(\ref{other})
 with $C_k = 1$ for all $k$.
Notice that on any lattice $C_k = 1$ for $k < r$, where $r$ is the
length of the smallest nontrivial loop on the lattice,
because the diagrams contributing to such $C_k$ can't
tell the difference between the given lattice and a
Bethe lattice of the same coordination number.

A stronger form of the positivity conjecture is that $C_k \geq 1$ for
bipartite lattices.

\subsection{Graphical expansion procedure for the Ising  model}
To make Section II more readable, we have confined into this subsection
some technical details on the graphical procedures used in the computation
of the Ising model HT expansions. 

 For simplicity, the whole graphical
expansion procedure can be split into three steps.  First, one has to
list all graphs entering into the calculation up to the maximum order
$L_{max}$ of expansion. To begin with, one forms the simple,
topologically distinct, one-vertex-irreducible graphs with $l \le
L_{max}$ edges.  One can further restrict to the subset of the
bipartite graphs, since only these can be embedded onto the bipartite
hsc or hbcc lattices.  This is the only memory intensive part of the
procedure, because there are many graphs\cite{bp2,bp3,bp4}
(approximately $3\cdot 10^5$ graphs at order 20, and over $5\cdot10^7$
at order 24), but it took only a few hours.  In a second step, the
lattice embedding-numbers and the symmetry-numbers of these graphs are
computed, one vertex of these graphs is ``marked'' in all possible
ways and the graphs are "decorated" to have also multiple lines.  This
is the subset of the graphs from which the expansion of the
magnetization can be reconstructed.

In the case of hsc lattices of high dimension, the most
time-consuming part of this procedure is the computation of the
embedding-number for each graph. In the case of the hbcc lattices the
timings are much smaller than for the hsc lattices and very slowly
dependent on $d$, but unfortunately the expansion coefficients are not
polynomials in $d$.  One begins by appropriately ordering the graph
vertices, and then the first of them is placed at the lattice
origin. The possible positions of the second vertex can be counted
exploiting the symmetries of the hyper-cube.  After fixing the first
two points of the embedding, the possible positions of the remaining
vertices are restricted to relatively few configurations by the
constraints given by the distances from the first two points and the
count can go on in a relatively easy way.  On the hsc lattices, the
timings for computing the magnetization expansion of the
$d$-dimensional Ising model at order $L_{max}$ increase exponentially
with the order of expansion and the lattice dimension $d$:
roughly as $O(5.5^{L_{max}} 2.5^d)$.  In particular, the computation
for the $10$-dimensional Ising model at order $20$ took $42$ days of
single-core time on a quad-core desktop computer with a CPU-clock
frequency of $2.8GHz$. Actually, less time was used since the
calculation was appropriately distributed on the four cores of the
computer.  Using more extensive computer resources, it would be
possible to compute only a few more orders, for not too high lattice
dimensions.

The next step implements the algebraic ``vertex-renormalization'',
namely the procedure of reconstruction\cite{wortis} of the
magnetization from the one-vertex-irreducible graphs having a single
marked vertex. By integrating the magnetization exactly with respect
to the field one finally obtains the free-energy in terms of the bare
vertices (up to a standard constant of integration).  This step of the
calculation is based on codes written in the Python language and is
fast.  The free-energy thus computed is model independent: eventually
one has to specialize the precise form of the bare vertex-functions to
the particular model of interest.

\section{Acknoledgements}
 We would like to thank David Bridges
for a helpful comment on the convergence of the expansion of the 
density-dependent dimer entropy.


\begin{thebibliography}{}
 \bibitem{10} M. E. Fisher, ``Statistical mechanics of dimers on a plane
 lattice", \emph{Phys. Rev.} {\bf 124}, 1664 (1961).
 
  \bibitem{11} P. W. Kasteleyn, ``The statistics of dimers on a lattice",
  \emph{Physica} {\bf 27}, 1209 (1961).

 \bibitem{19} R.J.Baxter, ``Dimers on a Rectangular Lattice", 
 \emph{J. Math. Phys.} {\bf 9}, 650 (1968).


\bibitem{fla} P. Flajolet and R. Sedgewick, {\it Analytic Combinatorics},
    Cambridge University Press (2009).

\bibitem{7} S. Friedland, E. Kropp, P.H. Lundow, K. Markstr\"{o}m,
``Validations of the Asymptotic Matching Conjectures",  \emph{J. Stat. Phys.} 
{\bf 133}, 513 (2008).

\bibitem{IQ2}A. Hanany and K. Kennaway, ``Dimer Models and Toric Diagrams", 
hep-th/0503149.

\bibitem{IQ3}R. Dijkgraaf, D. Orlando, and S. Reffert, ``Dimer Models, Free
Fermions and Super Quantum Mechanics",  \emph{Adv.Theor.Math.Phys.} {\bf 13},
1255 (2009).

\bibitem{ham}J. M. Hammersley,''Existence theorems and Monte Carlo
 methods for the monomer-dimer problem'' in {\it Research papers in
 statistics: Festschrift for J. Neyman}, edited by F.N. David. 
(Wiley, London 1966),  pag 125.

\bibitem{1} H. Minc, ``An Asymptotic Solution of the Multidimensional
Dimer Problem",  \emph{Lin.  Multilin. Alg.}, {\bf 8}, 235 (1980).

\bibitem{2} P. Federbush, ``Dimer $\lambda_d$ Expansion Computer
Computations", arXiv:math-ph/0804.4220v1.

\bibitem{3} P. Federbush, ``Dimer $\lambda_d$ Expansion, Dimension Dependence of $\bar{J}_n$ Kernels", arXiv:math-ph/0806.1941v1.

\bibitem{4} P. Federbush, ``Dimer $\lambda_d$ Expansion, A Contour Integral Stationary Point Argument", arXiv:math-ph/0806.4158v1.

  \bibitem{14} P. Federbush, ``Computation of Terms in the Asymptotic 
Expansion of Dimer $\lambda_d$ for High Dimensions'', 
\emph{Phys. Lett. A} {\bf 374}, 131 (2009).

\bibitem{6} S. Friedland and U.N. Peled, ``Theory of Computation of
Multidimensional Entropy with an Application to the Monomer-Dimer
Problem", \textit{Advances of Applied Math.} {\bf 34}, 486 (2005).


 \bibitem{13} P. Federbush and S. Friedland, ``An Asymptotic Expansion
 and Recursive Inequalities for the Monomer-Dimer Problem",
 \emph{J. Stat. Phys.} {\bf 143}, 306 (2011).

 \bibitem{17} P. Federbush, ``Convergence of the Formal 
Expansion for $\lambda_d(p)$ of the Monomer-Dimer Problem for 
Small $p$'', arXiv:1101.4591.

\bibitem{H-L} O.J.Heilmann and E.H.Lieb,``Theory of Monomer-Dimer
  Systems", Comm. Math. Phys. {\bf 25}, 190 (1972)

 \bibitem{ruelle}Ruelle, David, \emph{Statistical Mechanics}, 
W.~A. Benjamin, Inc. Amsterdam, 1969.
 

\bibitem{bp1} P. Butera and M. Pernici, ``Yang-Lee edge singularities
  from extended activity expansions of the dimer density for bipartite
  lattices of dimensionality $2 \leq d \leq 7$'', \emph{Phys. Rev.} 
E {\bf 86},  011104 (2012).

\bibitem{bp2} P. Butera and M. Pernici, ''Triviality problem and
  high-temperature expansions of higher susceptibilities for the Ising
  and scalar-field models in four-, five-, and six-dimensional
  lattices'', \emph{Phys. Rev.} E {\bf 85}, 021105 (2012).

\bibitem{bp3} P. Butera and M. Pernici, 
  ``High-temperature expansions of the higher susceptibilities for the
  Ising model in general dimension $d$'',  \emph{Phys. Rev.} E {\bf
    86}, 011139 (2012).

\bibitem{gaunt} D.S. Gaunt, ``Exact series-expansion study of the
  monomer-dimer problem'' , \emph{Phys. Rev.} {\bf 179}, 174 (1969).

\bibitem{kurfis}D. A. Kurtze and M. E. Fisher, ``Yang-Lee edge
  singularities at high temperatures'', \emph{Phys. Rev.} B {\bf 20},
  2785 (1979).

\bibitem{kenzie} S. McKenzie,  ``Extended high-temperature low-field
    expansions for the Ising model'', 
    \emph{Can. J. Phys.} {\bf 57}, 1239 (1979).

 
  \bibitem{12} P. Federbush, ``The Dimer Gas Mayer Series, the
 Monomer-Dimer $\lambda_d(p)$, the Federbush Relation,"
arXiv:1207.1252.
 
 \bibitem{15} P. Federbush, ``Asymptotic Expansions for $\lambda_d$ of 
the Dimer and Monomer-Dimer Problems", J. Stat. Phys. {\bf 150},  487 (2013).

\bibitem{rush} G.S. Rushbrooke, H.I. Scoins and A.J. Wakefield,
    ``The vapour pressures of athermal mixtures'', \emph{Discuss. Farad. Soc.}
    {\bf 15}, 57 (1953).

 \bibitem{domb} C. Domb, ``Ising model'',  in {\it Phase Transitions and Critical
Phenomena},  edited by C.~Domb and M. S. ~Green  
(Academic Press, New York 1974), vol. \ 3, pag 357.

\bibitem{wortis} M. Wortis, ``Linked cluster expansion''  in {\it Phase Transitions and Critical
Phenomena},  edited by C.~Domb and M. S. ~Green  
(Academic Press, New York 1974), vol. \ 3, pag 113.

\bibitem{fisga} M.E. Fisher and D.S. Gaunt, ``Ising Model and
  Self-Avoiding Walks on Hypercubical Lattices and High-Density
  Expansions'', \emph{Phys. Rev.} {\bf 133}, A224 (1964).

 \bibitem{20} P. Federbush, ``For the Monomer-Dimer $\lambda_d(p)$, 
the Master Algebraic Conjecture'', arXiv:1209-0987.

 
 
\bibitem{gutt} A. J. ~Guttmann, ``
 Asymptotic analysis of power-series expansions'',
 in "{\it Phase Transitions and Critical
Phenomena}", vol.\ 13, edited by C.~Domb and J.~Lebowitz 
(Academic Press, New York 1989),pag.1 .

\bibitem{beichl} I. Beichl, D.P. O'Leary and F. Sullivan,
  ``Approximating the number of monomer-dimer coverings in periodic
  lattices'', \emph{Phys. Rev.}  E {\bf 64}, 016701 (2001).

 \bibitem{21} P. Federbush, ``Dimer $\lambda_3=0.453(1) $  
and some other very intelligent guesses'', arXiv:0805.1195.
  

 \bibitem{schrij}A. Schrijver, ``Matching, edge-colouring, dimers'', in 
{\it Graph-theoretic concepts in computer science}, edited by H.L. Bodlaender
(Springer Lect. Notes in Computer Science 2880, Berlin 2003).

\bibitem{lund} P.H. Lundow, ``Compression of transfer matrices'',
    \emph{Discr. Math.} {\bf 231}, 321 (2001).

\bibitem{finch} S. R. Finch, {\it Mathematical Constants} (Cambridge
  University Press, Cambridge, 2003).
 

 \bibitem{Ftri} P.Federbush, ``For the Monomer-Dimer Problem on Triangular 
 and Hexagonal Lattices, the New $p$-Expansion",arXiv:1110.0684.

\bibitem{nagle} J.F. Nagle, ``New Series-Expansion Method for the Dimer 
    Problem'', \emph{Phys. Rev.} {\bf 152} 190 (1966).

\bibitem{stilck} J.F.Stilck and M.J. de Oliveira, ``Entropy of flexible
    chains placed on Bethe and Husimi lattices'', Phys. Rev. A {\bf 42},
    5955 (1990).  

\bibitem{bp4}P. Butera and M. Pernici,
 ``Free energy in a magnetic field and the universal scaling equation of state for the three-dimensional Ising model'', 
\emph{Phys. Rev.} B {\bf 83}, 054433 (2011).


\end{thebibliography}
\end{document}